\documentclass[runningheads]{llncs}

\usepackage{amssymb}
\usepackage{graphicx}

\usepackage{amsmath}
\usepackage{subfigure}

\usepackage{url}
\urldef{\mailsa}\path|{xixiluo.china}@gmail.com|
\urldef{\mailsb}\path|{shinaj}@rpi.edu|

\begin{document}
\mainmatter

\title{Simulating information creation in social Semantic Web applications}
\titlerunning{Simulating information creation in social Semantic Web applications}

\author{Xixi Luo\inst{1} \and Xiaowu Chen\inst{1} \and Qingping Zhao\inst{1}\and Joshua Shinavier\inst{2}}
\institute{State Key Laboratory of Virtual Reality Technology and Systems, Beihang University, Beijing 100191, China\\
\and Tetherless World Constellation, Rensselaer Polytechic Institute, Troy, NY 12180 USA}

\maketitle

\begin{abstract}
Appropriate ranking algorithms and incentive mechanisms are essential to the creation of high-quality information by users of a social network.  However, evaluating such mechanisms in a quantifiable way is a difficult problem.  Studies of live social networks of limited utility, due to the subjective nature of ranking and the lack of experimental control.  Simulation provides a valuable alternative: insofar as the simulation resembles the live social network, fielding a new algorithm within a simulated network can predict the effect it will have on the live network.  In this paper, we propose a simulation model based on the actor-concept-instance model of semantic social networks, then we evaluate the model against a number of common ranking algorithms.  We observe their effects on information creation in such a network, and we extend our results to the evaluation of generic ranking algorithms and incentive mechanisms.
\end{abstract}


\section{Introduction}

The Social Semantic Web \cite{conf:cssw2007}, \cite{book:socialSemanticWebKonvergenz} is a fairly new development that combines technologies, strategies and methodologies from the Semantic Web and social networks. It organizes its information by means of semi-formal ontologies, taxonomies or folksonomies, and it places a great deal of importance on community-driven semantics.  In the Social Semantic Web, the islands of the Social Web can be interconnected with semantic technologies, and Semantic Web applications are enhanced with the wealth of knowledge inherent in user-generated content \cite{citeulike:5398878}.

Since most of the information in a social network is contributed by online users, guiding users to create high-quality information is an important research topic.  This paper proposes a model to simulate information creation in Social Semantic Web applications. ``A simulation is an imitation of the operation of a real world process or system over time'' \cite{citeulike:763999}. Simulating information creation in semantic social networks can be used for the purpose of:
\begin{itemize}
\item predicting changes in the application. For example, the number of users in the application can be simulated. The likely effect of various courses of action can then be observed in the behavior of the model.
\item exploring the dynamics of the application.  Changing simulation settings and observing the result can provide valuable insight into the most important factors driving the evolution of the social network.
\item exploring new policies or mechanisms without disrupting ongoing operation of the real system. New policies or mechanisms can be tested without committing to a change in the actual social network.
\item studying Social Semantic Web applications in general.
\end{itemize}

Within the Semantic Web domain, simulations have been used for research into incentive mechanisms such as content trust\cite{DBLP:journals/ws/GilA07}.  However, these simulations are intended to validate specific incentive mechanisms. In contrast, the simulation framework proposed here can be used to evaluate general purpose incentive mechanisms.
This paper is a first step in exploring the simulation of information creation in social semantic web applications. The simulation model presented here is composed of \textit{actors} who carry out \textit{actions} in the application, and \textit{drivers}, or factors which affect information creation. Drivers can be classified as \textit{cost drivers} or \textit{reward drivers}, in that they determine the cost to an actor or the reward for an actor of carrying out a particular action. To simulate human being's instinctive reaction, only the reward of an action exceed the cost of the action will the actor carry it out in this simulation.

There are some researches about cost estimation model\cite{citeulike:5076084}\cite{boehm00:_cocomoII}, but so far we haven't find any reward estimation model which is combined together with the cost to simulate the execution of actions.

To demonstrate the simulation model, four ranking systems -- in-degree, PageRank, HITS and random ranking -- are tested in an experimental simulation.

We will first introduce the simulation model in greater detail, followed by four incentive mechanisms. We will then use the proposed simulation model to simulate the four ranking systems, and discuss the experimental results.

\section{Simulation Model}

Before we formally define the simulation model, we need to introduce some key concepts.
\begin{itemize}
\item A \textit{system} is ``a collection of entities (e.g. actors, concepts, and instances in this paper) that interact together over time to accomplish one or more goals'' \cite{citeulike:763999}.
\item A \textit{model} is ``an abstract representation of a system, usually containing structural, logical, or mathematical relationships that describe a system in terms of state, entities and their attributes, sets, processes, events, activities, and delay.'' \cite{citeulike:763999}
\end{itemize}

Building upon the above definitions, we will introduce the model's entities, attributes, activities, processes and states in turn.

\subsection{Entities}
An \textit{entity} is any object or component in the system that requires explicit representation in the model. The entities of this model are drawn from the actor-concept-instance model of ontologies \cite{mika:2006}, which contains the basic atom entities of an social semantic web application, the actor's participation of constructing concept and instance make the social semantic web social, and the concepts make the social semantic web distinct from other web applications from the semantic point of view. Without either one, we can't say it is a social semantic web application.
Entities are the subject of real-valued \textit{attributes} which are subject to various distributions.

In this model, a \textit {concept} (also known as a schema) is any tag, class, taxonomy or ontology which can be used to annotate or describe various data. The granularity of what we understand as a concept may vary widely, even within a single application.  For example, in Freebase\footnote{www.freebase.com}, we can consider both ``types'' and ``domains'' as concepts, depending on the requirements of the evaluation.

Furthermore, we associate with each concept a \textit{quality} attribute to indicate its rightness, completeness, ease of comprehension, and so on. The quality of a concept ranges from 0.0 (the lowest quality) to 1.0 (the highest quality). Note that in real systems, there is no such attributes like quality to be known, as it is not possible to ask each users for a quality for each concept. We propose the quality to estimate the average trust level of the concept from users.

Since this model is a open model, we can always personalize the model according to the system requirements by using different attributes and even more than one attributes. For example, we can set rightness or completeness and more as attributes instead of using an single attribute quality.

An \textit{instance} is the main carrier of information in this model. An instance can be a Web page, a photograph, an audio or video file, or any other object identifiable with a URI. Instances may be annotated with arbitrary concepts by \textit{actors}, or users. As with concepts, we use a ``quality'' attribute for instances which ranges from 0.0 to 1.0.

An \textit{annotation} is an actor-concept-instance tuple indicating that a particular actor has associated a particular instance with a particular concept.  In the following, if \begin{math}C\end{math} is a set of concepts, \begin{math}I\end{math} is a set of instances, and \begin{math}U\end{math} is a set of actors (users), then let \begin{math}A\subset U\times C\times I\end{math}  be the set of all possible annotations, with which actors in $U$ associate the instances of $I$ with concepts in $C$.

\subsection{The simulation process}
\begin{figure}
\centering
\includegraphics[scale=0.75]{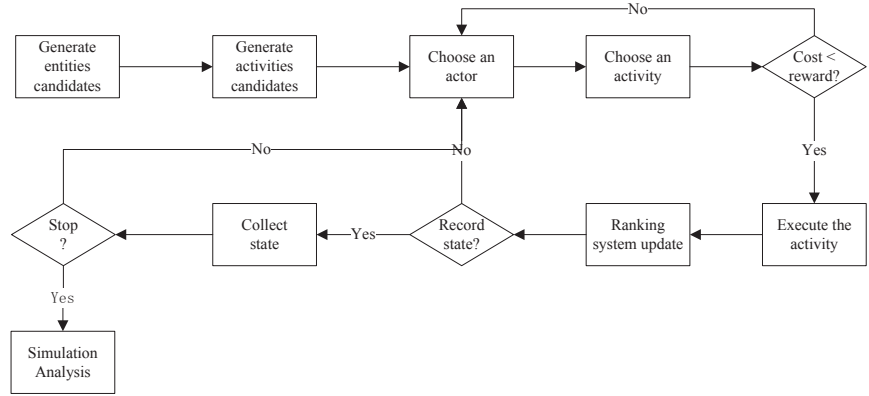}
\caption{the simulation process}
\label{fig:process}
\end{figure}

As illustrated in Figure \ref{fig:process}, the simulation process as a whole involves:
\begin{itemize}
\item the choice of an activity to carry out and actor to carry it out
\item the calculation of cost and reward of the action, or activity
\item the execution of the activity, if chosen, with corresponding effects on the simulation environment
\item the incremental ranking of entities
\item the optional recording of system state
\end{itemize}
Prior to simulation, candidate entities and activities for use in the simulation are generated.  A stop condition determines the end of the simulation, which is followed by simulation analysis.  We will introduce each of these events in turn.

\subsubsection{candidate generation}

The simulation starts with a preparatory phase in which entity and activity candidates are generated. The candidates' attributes follow specific distributions according to the requirements of the simulation. For example, in our experiment, actor candidates are associated with an ``expertise'' value which follows a normal distribution with a mean of 0.5 and a standard deviation of 0.5, the value above 1.0 or below 0.0 will be rounded up or down. Concept and instance candidates are associated with a ``quality'' value which follows a normal distribution with a mean of 0.5 and a standard deviation of 0.5. Activity candidates are distributed evenly among instance creation, concept creation, and semantic annotation.

\subsubsection{choice of actor and activity}
The body of the simulation consists of multiple iterations, each of which begins with the choice of an activity and an actor to carry it out. The actor is chosen randomly from among the actor candidates, and likewise, an activity is chosen randomly from among the activity candidates. Then the estimated cost and reward of the activity is calculated: if the cost is smaller than the reward, then the actor will carry out the activity.  Otherwise, the execution of the activity fails, and the simulation proceeds to the beginning of the next iteration. The calculation of the cost and reward of activities will be described in section \ref{section:activity customization}.

\subsubsection{activity execution}
If the chosen activity is found to be worthwhile (i.e. if the estimated reward exceeds the estimated cost), then it is carried out. The effects of the execution of the activity on the simulation environment vary according to the different activities.  For more details, see \ref{section:activities}.

\subsubsection{incremental ranking}
After an activity is carried out, the ranking of entities may need to be updated. This ranking is an important factor in the choice of the next activity, and it also influences the estimation of cost and reward. In general, the reward of an activity is higher if it involves a high-ranking entity, which reflects the greater visibility of the entity, in an application which features a recommendation system, and the greater inclination of users to choose it over less highly ranked entities.

Depending on the requirements of the simulation, the ranking can be updated after every iteration, or only occasionally.  To avoid excessive computational overhead, it should be possible to update the ranking incrementally, taking into account the changes which have occurred since it was last updated. We will discuss the details of the ranking system in the section of \ref{section_recommendation}.

\subsubsection{recording of system state}
To track the progress of the simulation, we need to record system state throughout the simulation process. In this experiment, we have chosen to record the extent of concept reuse, the quality of the most highly-ranked entities, and the rate of execution of potential activities. We will discuss the details of recording of system state in section of \ref{section:state}.

\subsubsection{stop condition}
The stop condition is when the time of successfully executed semantic annotation activities arrived a predefined number. For example, in this paper, the stop condition is 1000 times.

\subsection{activities}
\label{section:activities}
In this paper, we will describe seven types of activities: user registration, publishing of a concept, publishing of an instance, semantic annotation, linking of actors, linking of concepts, and linking of instances.
\begin{itemize}
\item \textit{user registration} is required in most social web applications, so that user activities can be tracked.
\item \textit{publishing} a concept or instance is analogous to publishing resources on the ordinary Web.  We are able to distinguish between two distinct types of resources -- concepts and instances -- and we consider the publishing of a concept and the publishing of an instance to be distinct activities.  Publishing an instance is the act of creating a web page, or uploading a photograph or an audio or video file to share with the community.  Publishing a concept, on the other hand, involves creating a tag, class, taxonomy or ontology which may be used to annotate instances. Since concepts may exist at different levels of granularity, even within the same application, we may consider more than one type of concept. In Freebase, for example, both domains and types are concepts, where a type is part of an domain: it is a finer-grained concept.
\item \textit{semantic annotation} distinguishes social semantic networks from most other Web applications. Actors associate instances with concepts to express a meaningful relationship. The concept's semantic value is used to organize or classify instances.
\item \textit{linking actors} makes a semantic social network social, in that the relationships between actors in a semantic social network comprise a social network. Linking actors is the activity of establishing a basic relationship between the actors, such as a ``friend'' relationship, or a ``knows'' relationship.
\item \textit{linking concepts} establishes semantic relationships among concepts. For example, simple hierarchical relationships are analogous to the sub- and superclass relationships of ontology languages such as OWL. Such linkage among concepts adds semantic value to instances annotated with those concepts.
\item \textit{linking instances} is similar to linking concepts. Examples include linking to a Web page from another Web page, linking a photograph to a set of photographs, and so on.
\end{itemize}

\subsection{Activity customization}\label{section:activity customization}
The detail of the execution of actions can be customized.
After customization, the reward and cost driver will be chosen and customized, which is the basis to calculate the estimated cost and reward during the simulation process. In this paper, the customization of the activities will be shown in the following:

\subsubsection{publishing concepts}
To publish a concept, the simulation randomly chooses a concept from the concept candidates, then the estimated cost and reward are calculated. If the reward exceeds the cost, then the concept is published. The estimated cost and reward are affected by the cost/reward drivers. The cost drivers of the ``publish concept'' activity are:
\begin{itemize}
\item CQ (concept quality), which is defined when the concept candidates are generated, whose values range from 0.0 to 1.0.
\item AE (the actor's expertise), which is defined when the actor candidates are generated, whose values range from 0.0 to 1.0.
\item CS (the concept's size), which is defined when the concept is generated, whose values range from 0.0 to 1.0.
\item AE\_PC (the actor's expertise in publishing concepts).  This is a cost driver which has a value of 1.0 when the actor has never published a concept, and 0.75 when the actor has only published one. In general, it is 1.0 divided by the number of concepts the actor has published.
\item UE\_PC (user effort for publishing a concept) is a cost driver which has a default value of 1.0. The higher the level is, the more it will cost to publish a concept.

\end{itemize}
We calculate the expected cost using the following formula, $\alpha$ is a prefixed parameter:
\begin{align}
cost = CS^\alpha \times CDs \\
CDs = (CQ + AE + AE\_PC) / 3\times UE\_PC
\end{align}

The reward drivers (whose values likewise range from 0.0 to 1.0) of publishing a concept action are:
\begin{itemize}
\item CQ (concept quality)
\item TCQ (the top concept's quality).  The top concept is the one with the highest reputation according to the recommendation system (see Section \ref{section_recommendation}). The reason we've chosen this as a reward driver is that the reward for an actor to publish a concept is lower if there already exists a highly ranked concept.
\item TCP (the top concept's popularity) is related to the total number of concepts. If the total number of published concepts is less than 10, for instance, or if the total number of published instances is less than 10, then the actor may still think he or she has a significant chance to create a very popular concept.  In this case, the value of TCP should be high. The greater the proportion of instances, among all instances, annotated by the best concept, the more ``dominant'' the concept is.  In this case, the expected reward is low: its value is is 1.0 minus the proportion just mentioned.
\end{itemize}
We calculate the expected reward by the following formula,$\beta$ is a prefixed parameter:
\begin{align}
reward = CQ^\beta\times RDs \\
RDs = (TCQ + TCP) / 2
\end{align}

\subsubsection{publishing instances}
Instances are chosen from the instance candidates, which are generated at the beginning of the simulation. The cost drivers for publishing instances include the instance type, instance size, and AE\_PI, the actor's expertise at publishing instances. Calculation of AE\_PI is similar to that of AE\_PC. Therefore, the formula used to calculate the expected cost is as follows (where UE\_PI is the user effort for publishing instances):
\begin{equation}
cost = AE\_PI \times UE\_PI
\end{equation}
To make things simpler, the reward drivers can be summarized as IQ (instance quality).
\begin{equation}
reward = IQ
\end{equation}

\subsubsection{semantic annotation}
In order to simulate the process of semantic annotation, one concept and one instance should first be chosen from the candidates, such that the chosen concept will be used to annotated the instance. In reality, a user would tend to choose the concepts that he is familiar with or that are easy to get.  In our simulation, we let the actor randomly choose a concept from his or her own concepts in addition to the top 10 concepts according to the recommendation system. If there is no recommendation system, then the actor will randomly choose one concept from his or her own concepts in addition to 10 random concepts. The procedure of choosing an instance is as follows: firstly, the actor will annotate his or her own instances until all instances have been annotated, at which point random instances are chosen.

The cost drivers of semantic annotation include:
\begin{itemize}
\item AE\_SA (the actor's expertise at semantic annotation), which is similar to AE\_PC and E\_PI
\item CC (the cost of choosing a concept), which is 0.0 if the concept is created by the actor but increases with ranking. For example, the cost of a concept in the top 10, is 0.1, while between the top 10 and top 20 it is 0.2 and so on. If the concept is not in the top 100, then the cost is 1.0.
\item CI (the cost of choosing an instance),which is calculated similarly to CC
\end{itemize}
The formula is as follows, where the UE\_SA is the level of user effort of semantic annotation:
\begin{equation}
cost = (AE\_SA + CC + CI )/3\times UE\_SA
\end{equation}

The reward drivers of semantic annotation include:
\begin{itemize}
\item CV (concept visibility). If the concept is the top 1 then the reward is very high: 1.0.  If it is only in the top 10, the reward is also high: 0.75.  In general, the reward decreasies with the rank: 1.0/(rank/10).
\item IV (instance visibility), which is calculated similarly to CV
\item CQ (the concept's quality)
\item IQ (the instance's quality)
\end{itemize}
The formula is as follows:
\begin{equation}
cost = ( CV + IV + CQ + IQ )/4
\end{equation}

\subsection{System state}
\label{section:state}
System state is a collection of variables that contain all the information necessary to describe the system at any time. The bellowing are the states we propose to record during the simulation.

\subsubsection{degree of concept reuse}
We use an entropy-based method to measure the degree of reuse of concepts.
If we were to simply use entropy to measure the uncertainty of concepts, the formula would be:
\begin{equation}H(X)=-\sum_i ^n p(c_i)\log p(c_i)\label{e3}\end{equation} where $p(c_i)=Pr(X = c_i)=\frac{|A_{c_i}|}{|A|}$ with  $|A_{c_i}|$ the number of annotations using concept $c_i$, and $|A|$ the total number of annotations.
For convenience, equation \ref{e3} may be expressed in the following as $H(X)=H(p(c_1), \dots, p(c_n))$ .

\begin{figure}
\label{f3}
\centering
\includegraphics[scale=0.45]{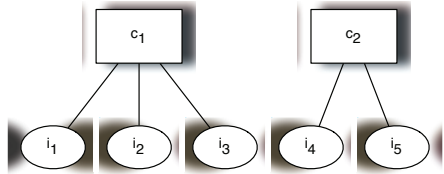}
\caption{multiple concepts without un-annotated instances}
\end{figure}

For example, if there is only one concept in an application, and all instances are associated with this concept, then $H(X)=-1\times \log 1 =0 $.  For the example in Figure\ref{f3}, $H(X)=-((\frac{3}{5}\times \log \frac{3}{5}) +(\frac{2}{5}\times \log \frac{2}{5}))=0.67$.

\begin{figure}
\subfigure[]
{
\label{f1a}
\includegraphics[scale=0.4]{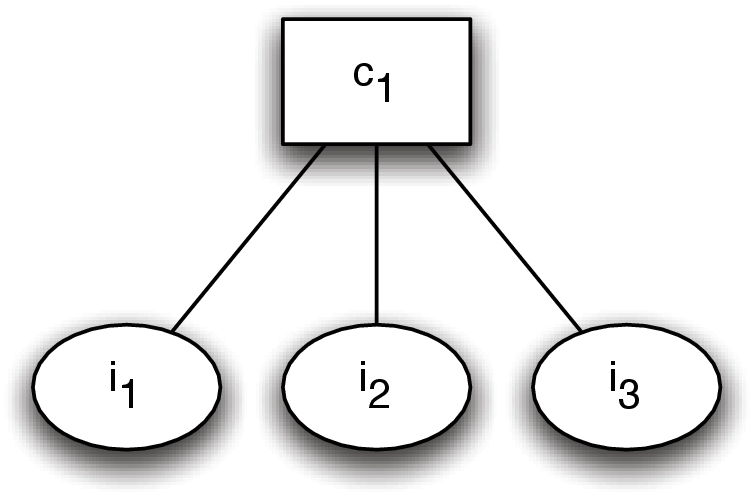}
}
\subfigure[]
{
\label{f1b}
\includegraphics[scale=0.4]{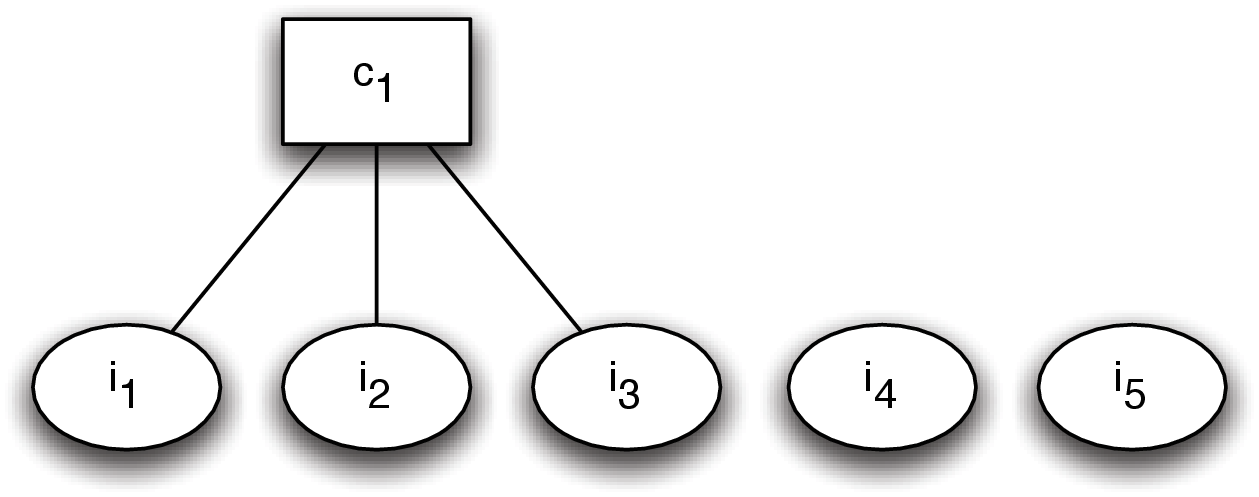}
}
\caption{Single concept}
\end{figure}

However, this simple metric falls short when applied to applications with instances which are not annotated. Consider the examples in Figure \ref{f1a} and \ref{f1b}. There are two un-annotated instances in Figure \ref{f1b}. According to Equation \ref{e3},  the value of $H$ for both examples should be 0, which is to say that all instances are annotated by the same concept. However, this is unintuitive for $i_4$ and $i_5$, which are not annotated at all.

Our solution to this problem is to import a virtual concept $c_v$ to the concept set $C$ to form a new set $C^*$, and then to annotate each of the un-annotated instances ``evenly'' by each concept. For example, if there are 99 concepts and one un-annotated instance, we will add one virtual concept for a total of 100 concepts, then for each concept, add $1/100^{th}$ of an annotation between the concept and instance.

\begin{figure}
\caption{single schema with un-annotated documents}
\label{fig:f2}
\centering
\includegraphics[scale=0.5]{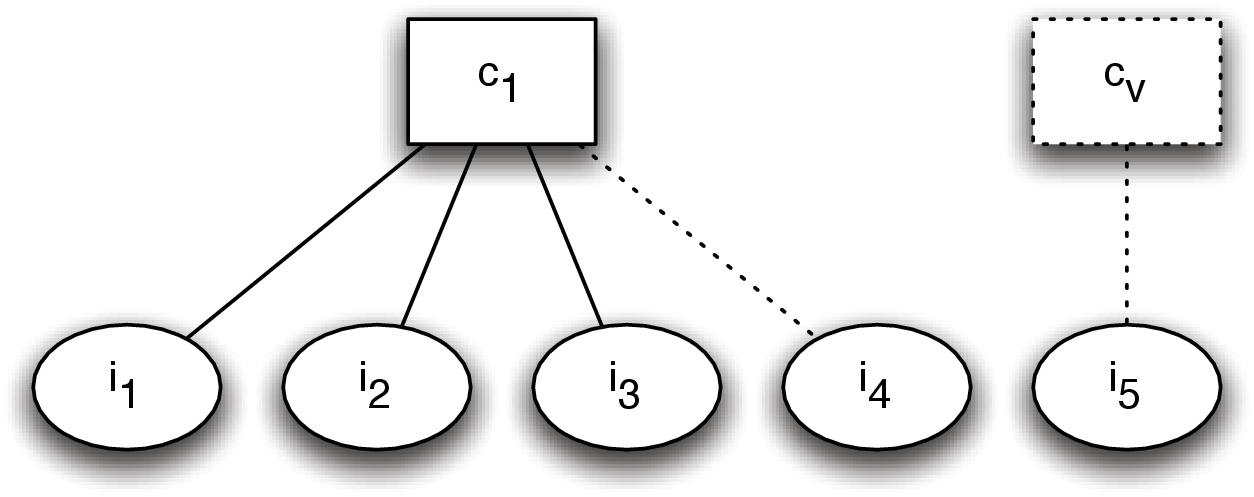}
\end{figure}

See Figure \ref{fig:f2}. After adding a virtual concept and distributing $i_4$ and $i_5$ to $c_1$ and $c_v$ respectively, the value of $H$ becomes $0.50$.

 For details about measuring degree of concept reuse, see \cite{DBLP:conf/aswc/LuoS09}.

\subsubsection{the quality of the top concepts}
We record the quality of the top entities to see how the overall quality of the top entities is affected by different ranking algorithms. Normally, the top entities are the most popular entities or the entities with top reputation provided by a recommendation system which ranks entities according to their reputation, using a specific algorithm.

\subsubsection{rate of execution of activities}
The rate of execution of activities is another attribute which we should track. We consider every iteration as a unit of time.  The total number of iterations can be considered as the total time of the simulation before it satisfies the stop requirement. The rate of execution of the semantic annotation activity is the number of successfully executed semantic annotations divided by the total number of semantic annotation activities that are chosen in the simulation. Since the stop condition is defined in terms of successfully executed semantic annotation activities, the execution rate of the semantic social network is an indication of how long the simulation takes. For example, if the execution rate of semantic annotation is $50\%$ and the stop condition is $1,000$ successful semantic annotation activities, then the total number of semantic annotations is $2,000$. Moreover, since the different activities are chosen randomly, the total semantic annotation is proportional to the total activities. Therefore, the execution rate of semantic annotation activities can be used to indicate how long the simulation takes. The greater the execution rate is, the less time the simulation takes, and vice versa.

\section{Ranking Algorithm}
\label{section_recommendation}
In this section, we will introduce four ranking algorithms which we are going to evaluate in this paper. These ranking algorithms will be applied to recommendation system, whose purpose is to guide users in their choice of one entity or another from the pool of published entities. For example, users choose between ontologies with which to annotate their instances, other users to add as friends a social network, or related instances to those they have created. The following ranking Algorithms are applied in order to rank ontologies, instances, and actors, providing ranked results to assist users in their decisions.
We assume the ranking algorithm is incremental, so at the end of each iteration of the simulation process, the ranking will be updated.

\subsection{Random ranking}
A random ranking is the baseline case.  By ``random'' we mean that there is no recommendation mechanism at all: users choose objects at will. For convenience, they will tend to choose the entities they are already familiar with, such as the entity they themselves have published. If the user hasn't published any resources, then they will randomly choose any entity from the pool of candidates.

%
%
%
%

\subsection {Indegree}

The indegree technique simply ranks nodes in a weighted, directed graph according to the total weight of edges directed at each node.  This is a very simple, and often effective, technique.

\subsection {Hits}

HITS\cite{citeulike:4156024}, or Hyperlink-Induced Topic Search, is a link analysis algorithm based on the notions of ``hubs'' and ``authorities'', which are defined in a mutual recursion.  Highly-ranked hubs are those nodes which link to highly-ranked authorities.  Highly-ranked authorities, in turn, are those nodes to which highly-ranked hubs link.  HITS typically operates over a defined subset of the overall network.

\subsection {PageRank}

The PageRank algorithm\cite{citeulike:3283}, like HITS, ranks nodes recursively according to the link structure of the network.  Unlike HITS, PageRank produces a global ranking of all nodes in the network.  It is very often used in search engines.  The MultiRank\cite{luo2009multirank} algorithm described in our previous work applies PageRank to the intermediate weighted graph described above.

\section{simulation results}

The simulation environment is set up as follows:
In the pool of objects, 100 actors, 1000 ontologies and instances, and 20,000 actions are generated. The 100 actors comprise a unchanging group of users, which is to say that we ignore the user registration process to make the simulation simpler to explain, since we want to focus the most important aspects of the model. The stop condition is that the semantic annotation action is successfully executed 1,000 times.
At each time step, an actor and an action are randomly chosen, then the estimated cost and reward are  calculated.  If the reward exceeds the cost, then the actor proceeds to execute the action.  Otherwise, the actor does nothing.
Every time an annotation action is successfully executed, we will record the entropy of the ontology in order to monitor ontology reuse, and also to keep track of the top popular otologies, in terms of quality. After the stop event occurs, we compute the execution rate of the semantic annotation actions for the purpose of estimating the execution time.

In this experiment, we only consider the publishing of concepts, the publishing of instances, and semantic annotation, and for each of these actions, we also consider the level of user effort during the calculation of estimated cost.

\subsection{user effort}
The level of user effort indicates the degree of effort required of the user of a particular application in order to complete a task. User effort varies by application. The value of the user effort level not limited to 0 and 1, since user effort is used to calculate the cost.  Low values for user effort, such as 0.1, mean that it cost less to execute the action, whereas higher values, such as 2.0, indicate higher cost. 1.0 is a default value of user effort, indicate the average level of user effort.

Here, we only consider the difference in user effort of semantic annotation. For the first group of experiments, user effort of semantic annotation has a value of 1.0.  For the second group, user effort of semantic annotation has a value of 2.0. We don't consider changing the level of user effort for publishing otologies or instances in this experiment, since in practice, the variance among application is not very high.  Therefore, we consider user effort of these actions to have a fixed value of 1.0.

The reason we set the level of semantic annotation user effort at 2.0 is that a lot of applications don't support semantic annotation very well.  We would like to see whether this will have a strong effect.

\subsection{analysis of results}

In this section, we will compare four simulation results, which include ontology entropy, the quality of the most-used otologies, and execution rate of semantic annotation.  The baseline is the random mechanism. The first group simulation setting is with semantic annotation support level 1.0, the second group simulation setting is with semantic annotation support level of 2.0. In this paragraph, we first discuss the four different results among different mechanism, and then compare the result with different semantic annotation support level.

After the execution session is the recording session, during which some statistical information is collected. The statistical information can be collected on every iteration, or only when some specific condition is satisfied. For example, the tracking of ontology reuse is conditional on a change in semantic annotations: we only calculate concept reuse when a semantic annotation activity occurs. What follows is a summary of the information we are going to collect at the end of each iteration.

\subsubsection{ontology entropy}
\begin{figure}
\subfigure[]
{
\label{fig:reuse-1}
\includegraphics[scale=0.45]{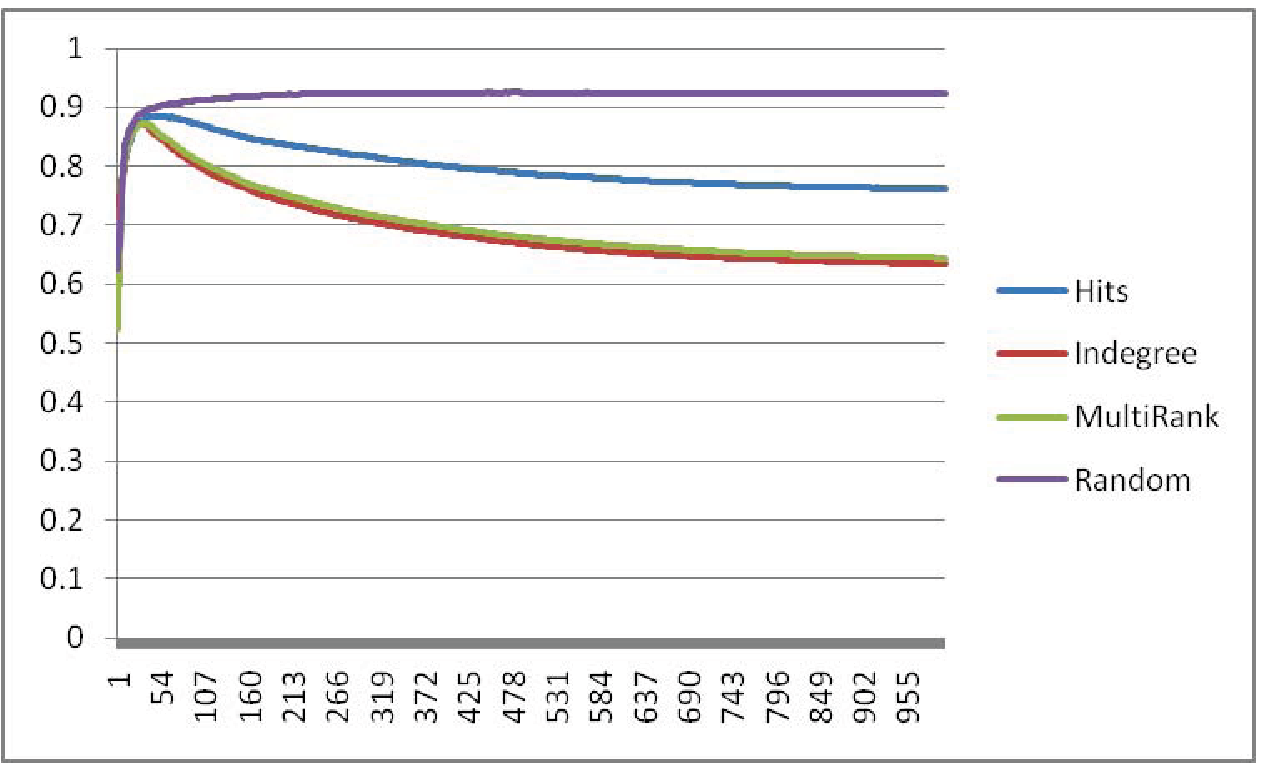}
}
\subfigure[]
{
\label{fig:reuse-2}
\includegraphics[scale=0.45]{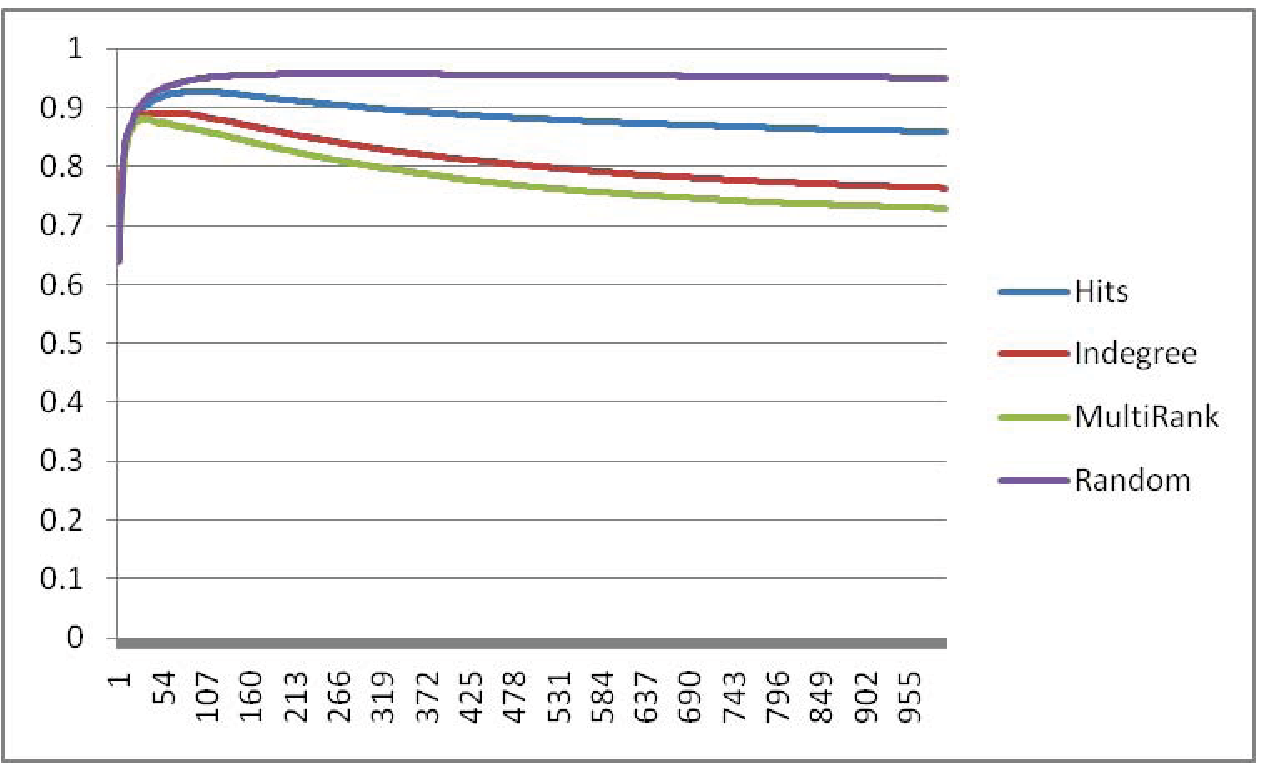}
}
\caption{entropy of concept reuse with SPL = 1.0 and 2.0}
\end{figure}
The figure\ref{fig:reuse-1} and \ref{fig:reuse-2} show that the random mechanism results in the highest entropy, followed by those based on the Hits, Indegree and PageRank algorithms.  This means that, in terms of promoting ontology reuse, PageRank is the most effective, in the case where the user has higher cost to semantic annotation (e.g. when the semantic annotation level is 2.0).

When the semantic annotation support level is 1.0, the best incentive mechanisms with respect to ontology reuse are PageRank and Indegree, followed by Hits, then Random. In semantic annotation, lower user effort is better for ontology reuse than higher effort.  For higher semantic annotation effort, the ontology entropy for PageRank and Indgree are between 0.7 and 0.8 (stable phase), while between 0.6 and 0.7 (stable phase) for lower effort. The ontology entropy for Hits also increases from between 0.7 and 0.8 to between 0.8 and 0.9 when the user effort of semantic annotation increases from 1.0 to 2.0. For the random mechanism, user effort makes little difference.

This can be explained by the fact that with low user effort for semantic annotation, users are encouraged to create annotations, increasing the likelihood of ontology reuse.

\subsubsection{quality of top ontologies}
\begin{figure}
\subfigure[]
{
\label{fig:top1concept-1}
\includegraphics[scale=0.45]{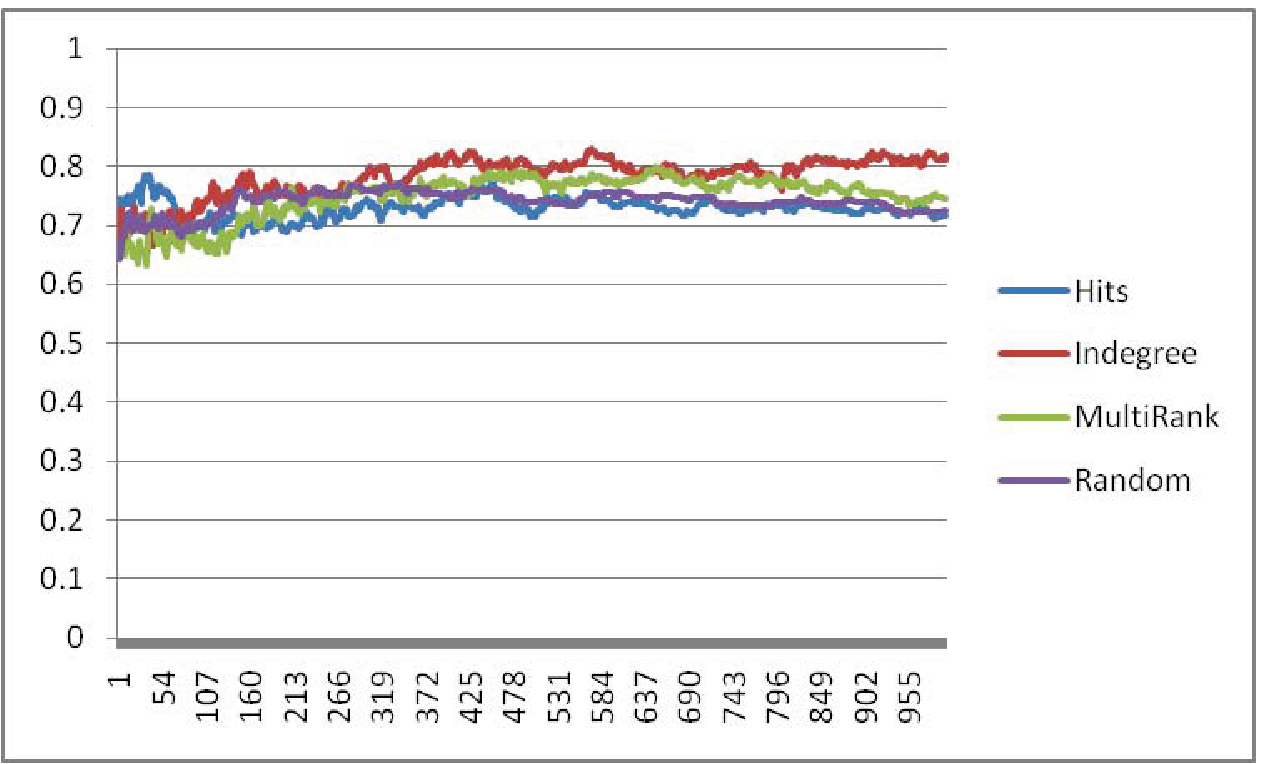}
}
\subfigure[]
{
\label{fig:top1concept-2}
\includegraphics[scale=0.45]{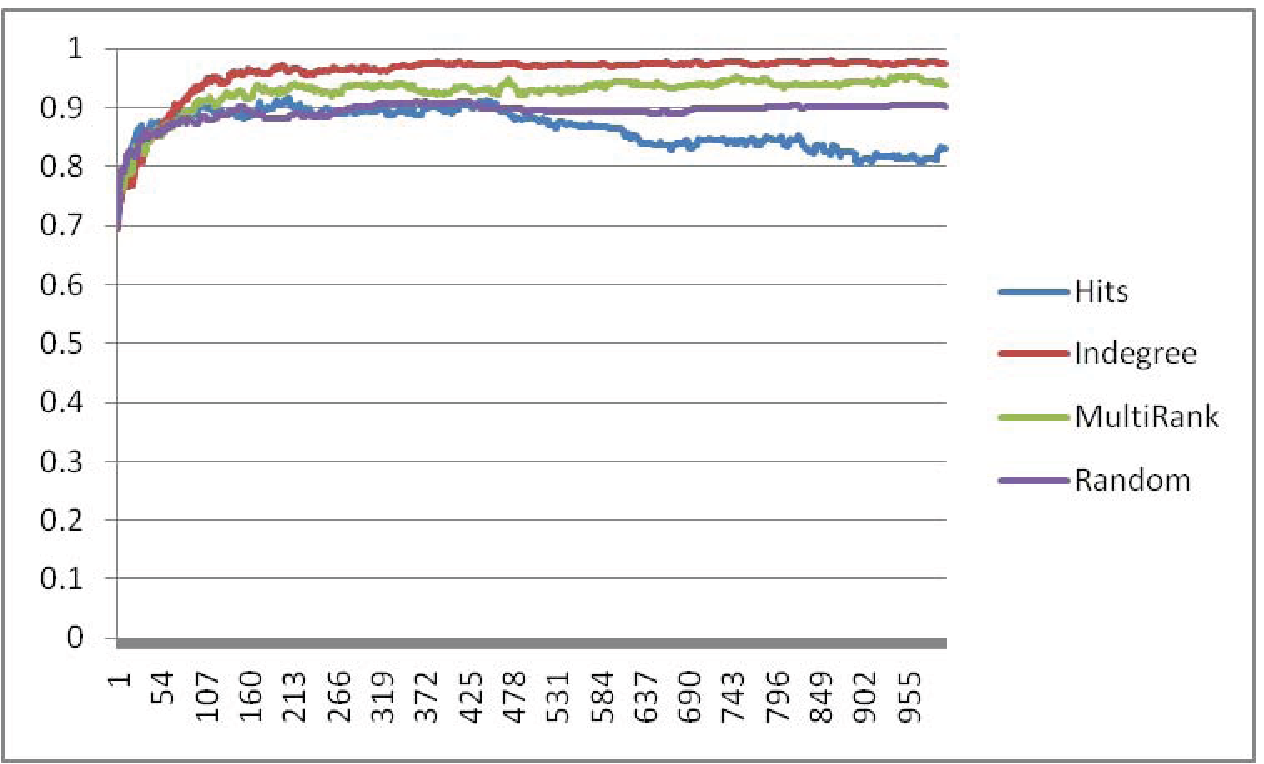}
}
\subfigure[]
{
\label{fig:top10concept-1}
\includegraphics[scale=0.45]{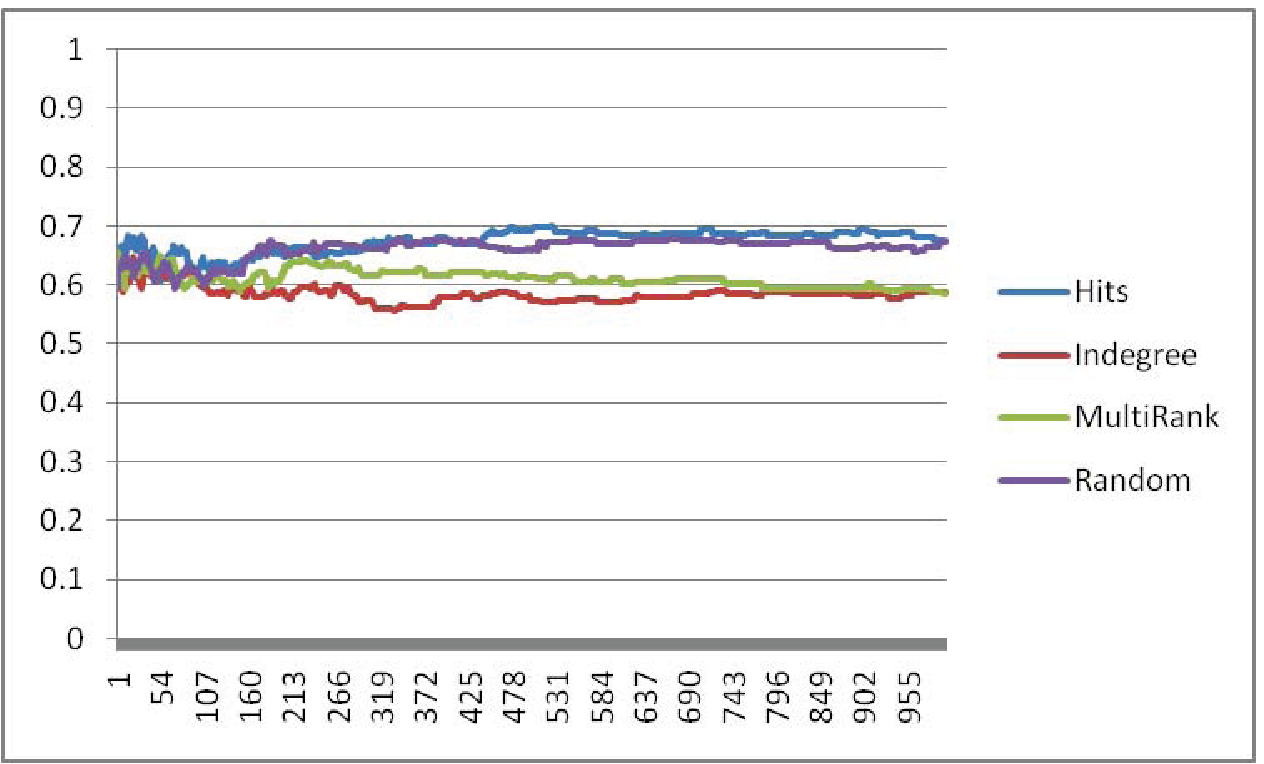}
}
\subfigure[]
{
\label{fig:top10concept-2}
\includegraphics[scale=0.45]{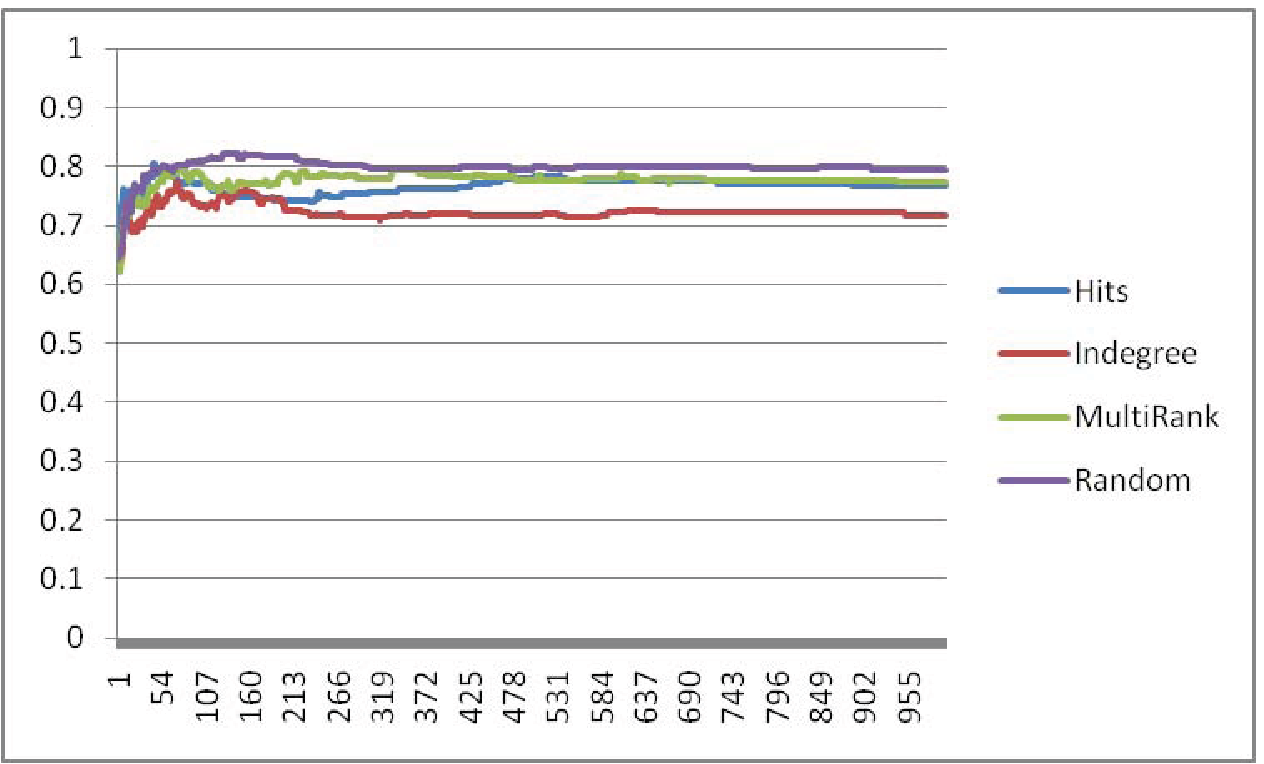}
}
\caption{top 1 and top 10 quality of concept with SPL = 1.0 and 2.0}
\end{figure}
The \textit{top} ontologies are the ones which have been reused the most. Reuse encompasses the activities of importing an existing ontology, and of using an ontology to annotate instances. In this experiment, we have only considered annotation, so the top ontology is the one which annotates the most instances.  After ranking ontologies accordingly, we find the quality of the top 1 ontology and the average quality of the top 10 ontologies.

Figures \ref{fig:top1concept-1}\ref{fig:top1concept-2}\ref{fig:top10concept-1}\ref{fig:top10concept-2} plot the quality of the top 1 and top 10 ontologies.
The column on the left (\ref{fig:top1concept-1} and \ref{fig:top10concept-1}) plots the quality resulting from a user effort level of 1.0, while the column on the right (\ref{fig:top1concept-2}\ref{fig:top10concept-2}) plots the quality resulting from a user effort level of 2.0. We can see that higher user effort results in higher quality. For the top 1 ontology, when the user effort level increases from 1.0 to 2.0, the quality increases correspondingly from a value between 0.7 and 0.8 to a value between 0.8 and 1.0. In the case of the top 10 ontologies, when the user effort level increases from 1.0 to 2.0, the quality increases from a value between 0.6 and 0.7 to a value between 0.7 and 0.8, again confirming the positive effect of user effort on quality.

The top row (\ref{fig:top1concept-1} and \ref{fig:top1concept-2}) plots the quality of the top 1 ontology, while the bottom row (\ref{fig:top10concept-1}\ref{fig:top10concept-2}) plots the quality of the top 10 ontologies, respectively, with user effort level of 1.0 and 2.0. We found significant differences between the top 1 and top 10 cases, in terms of quality. For example, the indegree mechanism results in the highest quality in the top 1 case, but the lowest quality in the top 10 case, which means that the top 1 ontology has a much higher quality than that of the other ontologies in the top 10.  That is to say, the use of Indegree causes top ontologies to dominate.  This is valuable from the point of view of promoting ontology reuse.

When the user effort of annotation is 2.0 (meaning that relatively high effort is required to annotate instances), the quality of the top 1 and top 10 ontologies is as depicted in the figure \ref{fig:top1concept-2}and \ref{fig:top10concept-2}. For the top 1 ontology, Indegree yields the highest average quality, followed by MultiRank, Random and Hits. The quality produced by Indegree, MultiRank, and Random starts between 0.6 and 0.7 (which is the average quality of all the ontologies), and then increases to above 0.9 after 200 iterations, while Hits decrease to below 0.9 (still above 0.8) after around 400 iterations. Here, the Random mechanism seems to do better than Hits, which seems counterintuitive. However, it does make sense that the greater the user effort required for semantic annotation, the less important the ranking mechanism is.  As a result it becomes increasingly unlikely that the user will execute the action.  Does this mean that less supportive environments actually facilitate ontology reuse?  We will see in the following section that this is not the case.

\subsubsection{rate of semantic annotation}
The rate of creation of semantic annotations is an important factor for estimating the length of the rest of the simulation. The stop condition of the simulation is a specific number of successfully executed semantic annotation actions, therefore the execution rate of semantic annotation can be used to estimate how long will it take to arrive at this predefined number of actions. The lower the semantic annotation rate, the longer the simulation takes, which corresponds to the reality that if users don't have a high rate of semantic annotation, then the system takes a long time to accumulate a specific amount of annotation information.
\begin{figure}
\subfigure[]
{
\label{fig:saRate-1}
\includegraphics[scale=0.45]{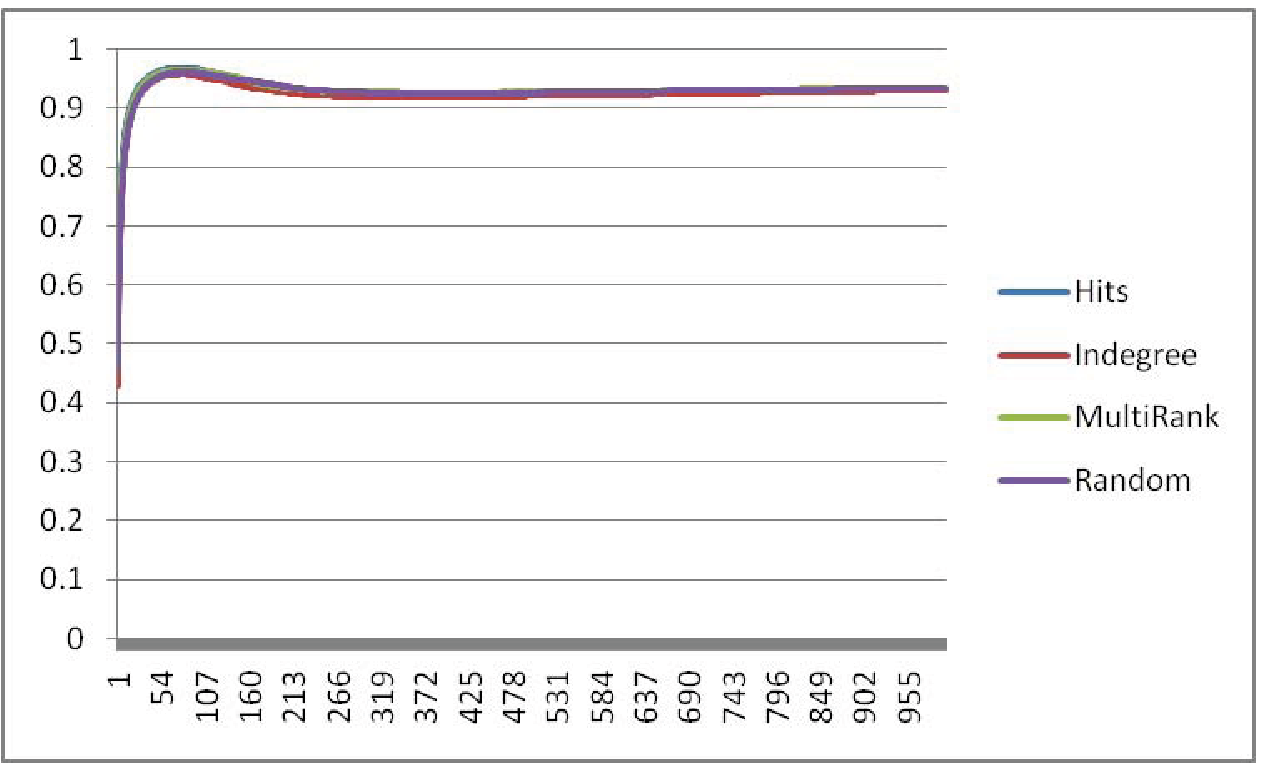}
}
\subfigure[]
{
\label{fig:saRate-2}
\includegraphics[scale=0.45]{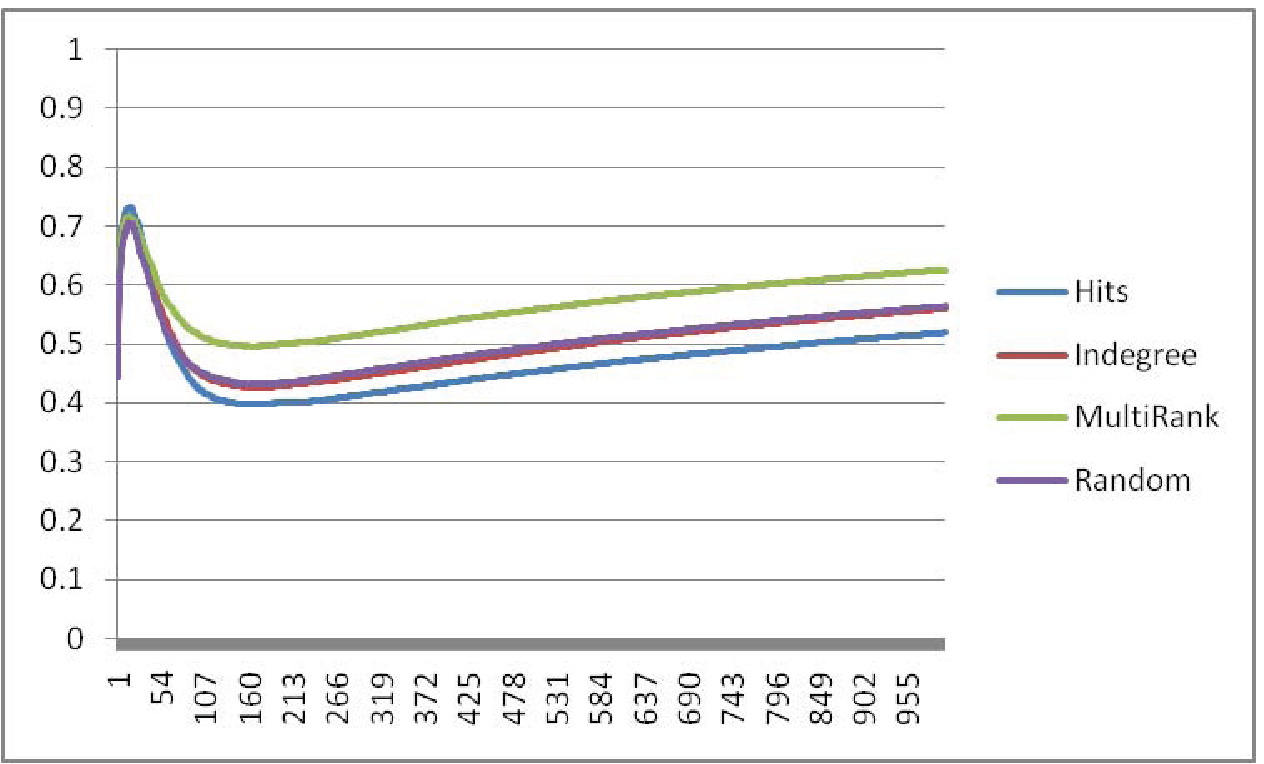}
}
\caption{Semantic Annotation Rate with SPL = 1.0 and 2.0}
\end{figure}
The \ref{fig:saRate-1} is the semantic annotation rate when SPL is 1.0, and the \ref{fig:saRate-2} is the semantic annotation rate when SPL is 2.0. When it cost less (i.e. when SPL is 1.0) to perform semantic annotation, there is no difference between the four mechanisms. When it cost more (i.e. when SPL is 2.0), MultiRank takes the least time, followed by Indegree and Random, and then Hits, which is to say that the if the system uses the MultiRank ranking system, it will take less time to arrive at the predefined number of semantic annotations.

\section{Conclusion and Future Work}
In this paper, we have illustrated the use of a simulation to evaluate ranking algorithms and incentive mechanisms in a quantifiable way.  Using this technique, different mechanisms can be tested in advance of their deployment in a live social network environment.  We have made two assumptions:
\begin{itemize}
\item it is possible to mimic the behavior of a real social network by means of a simulation
\item the simulation model described above is appropriate for modeling information creation in semantic social networks
\end{itemize}
In the past, ranking algorithms have tended to be evaluated only subjectively: if the algorithm produces subjectively accurate results, then it is an appropriate algorithm.  The technique presented in this paper provides a measurable alternative.  However, the question naturally arises of whether this technique itself is appropriate.  How do we evaluate it?  Until such time as there are formal techniques for evaluating social network simulations, we must rely on subjective evaluation, choosing the simulation model as sensibly as possible and allowing the results to speak for themselves.  We have thoroughly described our simulation model and justified these choices wherever possible: our object model is based on the actor-concept-instance model of social-semantic networks, where change in the network is driven by an iterative process of user actions.  Users are guided by metrics of cost, reward in conjunction with predefined metrics of quality as well as the ranking algorithms under investigation.  We have presented the results of applying our technique to several common ranking algorithms, and we have shown that these results are reasonable.

In future, we intend to test our simulation model against multi-relational ranking algorithms which take advantage of the ``semantics'' of social-semantic networks.  We plan to release our implementation software under an open-source license, so that it can be freely reused by developers of social networking applications.

\section{Acknowledgements}

We would like to thank Jim Hendler and Deborah McGuinness for their valuable feedback on various drafts of this paper.

\bibliographystyle{amsplain}
\bibliography{simulation}
\end{document}